\documentclass[11pt]{article}
\usepackage{amsmath,epsfig}
\begin{document}

\renewcommand{\theequation}{\thesection.\arabic{equation}}
\newcounter{saveeqn}
\newcommand{\add}{\addtocounter{equation}{1}}
\newcommand{\alpheqn}{\setcounter{saveeqn}{\value{equation}}%
\setcounter{equation}{0}%
\renewcommand{\theequation}{\mbox{\thesection.\arabic{saveeqn}{\alph{equation}}}}}
\newcommand{\reseteqn}{\setcounter{equation}{\value{saveeqn}}%
\renewcommand{\theequation}{\thesection.\arabic{equation}}}
\newenvironment{nedalph}{\add\alpheqn\begin{eqnarray}}{\end{eqnarray}\reseteqn}
\newsavebox{\PSLASH}
\sbox{\PSLASH}{$p$\hspace{-1.8mm}/}
\newcommand{\PS}{\usebox{\PSLASH}}
\newsavebox{\PARTIALSLASH}
\sbox{\PARTIALSLASH}{$\partial$\hspace{-2.3mm}/}
\newcommand{\PARTIALS}{\usebox{\PARTIALSLASH}}
\newsavebox{\sSLASH}
\sbox{\sSLASH}{$s$\hspace{-1.9mm}/}
\newcommand{\sS}{\usebox{\sSLASH}}
\newsavebox{\KSLASH}
\sbox{\KSLASH}{$k$\hspace{-1.8mm}/}
\newcommand{\KS}{\usebox{\KSLASH}}
\newsavebox{\LSLASH}
\sbox{\LSLASH}{$\ell$\hspace{-1.8mm}/}
\newcommand{\LS}{\usebox{\LSLASH}}
\newsavebox{\QSLASH}
\sbox{\QSLASH}{$q$\hspace{-1.8mm}/}
\newcommand{\QS}{\usebox{\QSLASH}}
\newsavebox{\DSLASH}
\sbox{\DSLASH}{$D$\hspace{-2.5mm}/}
\newcommand{\DS}{\usebox{\DSLASH}}
\newsavebox{\DbfSLASH}
\sbox{\DbfSLASH}{${\mathbf D}$\hspace{-2.8mm}/}
\newcommand{\DBFS}{\usebox{\DbfSLASH}}
\newsavebox{\DELVECRIGHT}
\sbox{\DELVECRIGHT}{$\stackrel{\rightarrow}{\partial}$}
\newcommand{\PARVECR}{\usebox{\DELVECRIGHT}}

\title{\bf Semi-Leptonic Decay of a Polarized Top Quark in the Noncommutative Standard Model}
\author{ \bf M. Mohammadi
Najafabadi\thanks{email: mojtaba.mohammadi.najafabadi@cern.ch, Now at CERN}\\ 
\\
{\it Department of Physics, Sharif University of Technology}\\
{\it P.O. Box 11365-9161, Tehran-Iran} \\
 and\\
{\it Institute for Studies in Theoretical Physics and Mathematics (IPM)}\\
{\it School of Physics, P.O. Box 19395-5531, Tehran-Iran} \\}

\date{}
\maketitle

\begin{abstract}
In this paper we study the noncommutative effects to the lepton
spectrum from the decay of a polarized top quark. It is shown that
the lowest contribution comes from the quadratic terms of the
noncommutative parameter. The deviations from the standard model are
significant for small values of the noncommutative characteristic
scale. However, the charged lepton spin correlation coefficient
has a remarkable deviation from the standard model from very
low values of the noncommutative characteristic scale to 1 TeV.
\end{abstract}

\vspace{4cm}
\hspace{0.8cm}

\newpage
\setcounter{page}{1}
\section{Introduction}
\setcounter{section}{1} The standard model of the particles has been
found to be in agreement with experiment in many of its aspects.
However, in the framework of the standard model top is the only
quark which has a mass in the same order as the electroweak symmetry
breaking scale, $v\sim 246$ GeV, whereas all other observed fermions
have masses which are a tiny fraction of this scale. This huge mass
might be a hint that top quark plays an essential role in the
electroweak symmetry breaking. On the other hand, the reported
experimental data from Tevatron on the top quark properties are
still limited and no significant deviations from the standard model
predictions has been seen \cite{Tevatron}. The number of observed
top quark events in the Tevatron experiment is increasing and now
reaching to the order of a few hundred. Several properties of the
top quark have been already examined. They consist of studies of the
$t\bar{t}$ production cross section, the top quark mass measurement,
the measurement of $W$ helicity in the top decay, the search for
FCNC and many other studies \cite{Tevatron}. However, it is expected
that top quark properties can be examined with high precision at the
LHC due to very large statistics \cite{Beneke}. Since the dominant
top quark decay mode is into a $W$ boson and a bottom quark, the
$tWb$ coupling can be investigated accurately. Within the standard
model, the top quark decay via electroweak interaction before
hadronization. This important property is one the consequences of
its large mass. Hence, the spin information of the top quark is
transferred to its decay daughters. Thus, the top quark spin can be
used as a powerful mean for investigation of any possible new
physics.

There are many studies for testing the top quark decay properties at
hadron colliders. For instance, the non-standard effects on the full
top width have been investigated in the minimal supersymmetric
standard model and in the technicolor model \cite{BSM}. Some studies
have been performed on the effects of anomalous $tWb$ couplings on
the top width and some constraints have been applied on the
anomalous couplings \cite{ATOP}. There have been some studies on the
top quark rare decays. The two-body decay of the top quark,
$t\rightarrow Wb$, has been considered within the noncommutative
standard model in \cite{Namit}.

In this paper we study the semi-leptonic decay of the top quark,
$t\rightarrow l\nu b$, in the noncommutative standard model. The
noncommutativity in space-time is a possible generalization of the
usual quantum mechanics and quantum field theory to describe the
physics at very short distances of the order of the Planck length,
since the nature of the space-time changes at these distances.
 The noncommutative spaces can be realized as spaces where coordinate operators,
$\hat{x}_{\mu}$, satisfy the commutation relations:
\begin{eqnarray}
[\hat{x}_{\mu},\hat{x}_{\nu}] = i\theta_{\mu \nu},
\end{eqnarray}
where $\theta_{\mu \nu}$ is a real antisymmetric tensor with the
dimension of $[L]^{2}$. We note that a space-time noncommutativity,
$\theta_{0i}\neq 0$, might lead to some problems with unitarity and
causality \cite{Gomis}. A noncommutative version of an ordinary
field theory can be obtained by replacing all ordinary products with
Moyal $\star$ product defined as \cite{review}:
\begin{eqnarray}
(f\star g)(x) &=& \exp\left(\frac{i}{2}\theta^{\mu
\nu}\partial_{\mu}^{y}\partial_{\nu}^{z}\right)f(y)g(z)\bigg\vert_{y=z=x}\\
\nonumber &=&
f(x)g(x)+\frac{i}{2}\theta^{\mu\nu}(\partial_{\mu}f(x))(\partial_{\nu}g(x))+O(\theta^{2}).
\end{eqnarray}
The approach to the noncommutative field theory based on the Moyal
product and Seiberg-Witten maps allows the generalization of the
standard model to the case of noncommutative space-time, keeping the
original gauge group and particle content \cite{SW,Madore}.
Seiberg-Witten maps relate the noncommutative gauge fields and
ordinary fields in commutative theory via a power series expansion
in $\theta$. Indeed the noncommutative version of the standard model
is a Lorentz violating theory, but the Seiberg Witten map shows that
the zeroth order of the theory is the lorentz invariant standard
model. The effects of the noncommutative space-time on some rare and
collider processes have been considered in \cite{Haghighat}. The
leptonic decay of the $W$ and $Z$ bosons on the noncommutative
space-time have been studied in \cite{Iltan}.

This paper is organized as follows: In Section 2, a short introduction
for the NCSM is given. Section 3 is dedicated to present the noncommutative effects
on the $t(\uparrow)\rightarrow l^{+}+\nu_{l}+b$ decay. Finally, Section 4 concludes 
the paper.

\section{Noncommutative Standard Model (NCSM)}
The action of the NCSM can be obtained by replacing the ordinary
products in the action of the classical SM by the Moyal products and
then matter and gauge fields are replaced by the appropriate
Seiberg-Witten expansions. The action of NCSM can be written as:
\begin{eqnarray}
S_{NCSM} = S_{fermions} + S_{gauge} + S_{Higgs} + S_{Yukawa},
\end{eqnarray}
We just consider the fermions (quarks and leptons). The fermionic
matter part in a very compact way is:
\begin{eqnarray}
S_{fermions} = \int d^{4}x
\sum_{i=1}^{3}\left(\bar{\widehat{\Psi}}^{(i)}_L\star
(i\widehat{\DS} ~\widehat{\Psi}^{(i)}_L)\right) + \int d^{4}x
\sum_{i=1}^{3}\left(\bar{\widehat{\Psi}}^{(i)}_R\star
(i\widehat{\DS} ~\widehat{\Psi}^{(i)}_R)\right),
\end{eqnarray}
where $i$ is generation index and $\Psi^{i}_{L,R}$ are:
\begin{eqnarray}
\Psi^{(i)}_L = \left(
                 \begin{array}{c}
                   L^{i}_{L} \\
                   Q^{i}_{L} \\
                 \end{array}
               \right)
~,~\Psi^{(i)}_R = \left(
                 \begin{array}{c}
                   e^{i}_{R} \\
                   u^{i}_{R} \\
                   d^{i}_{R}
                 \end{array}
               \right)
\end{eqnarray}
where $L^{i}_{L}$ and $Q^{i}_{L}$ are the well-known lepton and
quark doublets, respectively. The Seiberg-Witten maps for the
noncommutative fermion and vector fields yield:
\begin{eqnarray}
\widehat{\psi} &=& \widehat{\psi}[V] = \psi -
\frac{1}{2}\theta^{\mu\nu}V_{\mu}\partial_{\nu}\psi +
\frac{i}{8}\theta^{\mu\nu}[V_{\mu},V_{\nu}]\psi + O(\theta^{2}),\nonumber \\
\widehat{V_{\alpha}} &=& \widehat{V_{\alpha}}[V] = V_{\alpha} +
\frac{1}{4}\theta^{\mu\nu} \{\partial_{\mu}V_{\alpha} +
F_{\mu\alpha}, V_{\nu}\} + O(\theta^{2}),
\end{eqnarray}
where $\psi$ and $V_{\mu}$ are ordinary fermion and gauge fields,
respectively. Noncommutative fields are denoted by a hat. For a full
description and review of the NCSM, see \cite{Madore}. The
$t(p_{1})\rightarrow W(q)+b(p_{2})$ vertex in the NCSM up to the
order of $\theta^{2}$ can be written as \cite{Namit,Madore}:
\begin{eqnarray}\label{vertex}
\Gamma_{\mu,NC} &=& \frac{g V_{tb}}{\sqrt{2}}[\gamma_{\mu} +
\frac{1}{2}(\theta_{\mu\nu}\gamma_{\alpha}+\theta_{\alpha\mu}\gamma_{\nu}+
\theta_{\nu\alpha}\gamma_{\mu})q^{\nu}p^{\alpha}_{1} \\
\nonumber
&-&\frac{i}{8}(\theta_{\mu\nu}\gamma_{\alpha}+\theta_{\alpha\mu}\gamma_{\nu}+
\theta_{\nu\alpha}\gamma_{\mu})(q\theta
p_{1})q^{\alpha}p^{\nu}_{1}]P_{L}.
\end{eqnarray}
where $P_{L} = \frac{1-\gamma_{5}}{2}$ and $q\theta p_{1}\equiv
q^{\mu}\theta_{\mu\nu}p^{\nu}_{1}$. This vertex is similar to the
vertex of $W$ decays into a lepton and anti-neutrino
\cite{Madore,Iltan}.

\section{The noncommutative effects on the $t(\uparrow)\rightarrow l^{+}+\nu_{l}+b$ decay}
\begin{figure}[ht]
\begin{center}
\epsfig{file=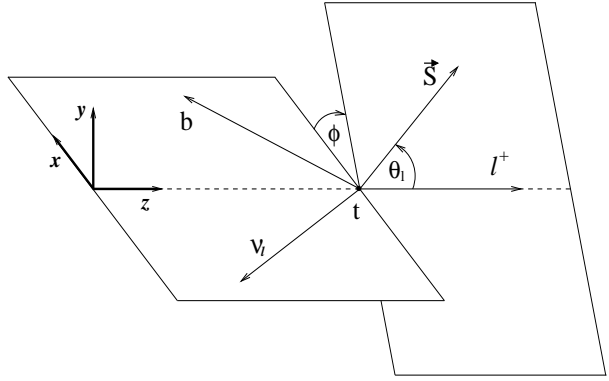,height=5cm,width=8cm}
 \caption{The process $t(\uparrow)\rightarrow l^{+}+\nu_{l}+b$ in the rest frame of the top quark.}
    \label{EventPlane}
\end{center}
\end{figure}

The effective vertex in Eq.(\ref{vertex}) contains $\gamma_{\mu}$
and the momenta of the involved particles. In order to simplify the
calculations we ignore of the masses of the $b$ quark and leptons.
By using the Dirac equations, the vertices have simpler forms.
For instance, the Eq.(\ref{vertex}) can be replaced by the following:
\begin{eqnarray}
&&\bar{u}(p_{2})\Gamma^{\mu}u(p_{1}) = \left(\frac{g}{\sqrt{2}}\right)V_{tb}\times \nonumber \\
&&\bar{u}(p_{2})\left((1+\frac{1}{2}q\theta
p_{1}+\frac{i}{8}(q\theta
p_{1})^{2})\gamma^{\mu}P_{L}-m_{t}(\frac{1}{2}+\frac{i}{8}q\theta
p_{1})\theta^{\mu\nu}p_{2,\nu}P_{R}\right)u(p_{1}),
\end{eqnarray}
According to the Fig.\ref{EventPlane}, the event plane defines the
(x-z) plane. The z-axis is determined by the momentum of the lepton
and $\theta_{l}$ is the angle between momentum of the lepton and the
spin of the top quark in the rest frame of the top quark. Using the
following identities:
\begin{eqnarray}
u_{\alpha}\theta^{\alpha\beta}v_{\beta} = u\theta v =
\vec{\theta}.(\vec{u}\times
\vec{v})~,~u_{\mu}\theta^{\mu\nu}\theta^{\alpha}_{\nu}v_{\alpha} =
|\vec{\theta}|^2(\vec{u}.\vec{v}) -
(\vec{u}.\vec{\theta})(\vec{v}.\vec{\theta}),
\end{eqnarray}
where $\vec{\theta} = (\theta_{23},\theta_{31},\theta_{12})$ 
and if we assume that $\vec{\theta}$
is in the (x-z) plane the squared matrix element for the 
reaction $ t(p_{1})\rightarrow W(q) + b(p_{2}) \rightarrow
l^{+}(k_{1})+\nu_{l}(k_{2})+b(p_{2})$ has the following form:
\begin{eqnarray}
\overline{|\mathcal{M}|}^{2} \propto
\left((k_{1}.\hat{p}_{1})(k_{2}.p_{2}) + 
\frac{m_{t}^{2}}{16}(|\overrightarrow{\bf p_{2}}|^{2}|\overrightarrow{\bf \theta}|^{2}-
( \overrightarrow{\bf p_{2}}.\overrightarrow{\theta})^{2})(k_{1}.k_{2})(\hat{p}_{1}.p_{2})\right),
\end{eqnarray}
where $\hat{p}^{\mu}_{1} = p^{\mu}_{1} -m_{t}s^{\mu}$ and $s^{\mu} =
(0,\vec{s})$ is the polarization four-vector of the top quark.

The differential rate for $t\rightarrow l^{+}+\nu_{l}+b$ at Born approximation can be written as:
\begin{eqnarray}\label{MX}
&&d\Gamma=\frac{1}{2m_{t}}\frac{64G^{2}_{F}}{(1-y/r)^{2}+\gamma^{2}_{W}}\frac{dR_{3}}{(2\pi)^{5}}\times\nonumber\\
&&\left((k_{1}.\hat{p}_{1})(k_{2}.p_{2}) + 
\frac{m_{t}^{2}}{16}(|\overrightarrow{\bf p_{2}}|^{2}|\overrightarrow{\bf \theta}|^{2}-
( \overrightarrow{\bf p_{2}}.\overrightarrow{\theta})^{2})(k_{1}.k_{2})(\hat{p}_{1}.p_{2})\right),
\end{eqnarray}
where $r=m^{2}_{W}/m^{2}_{t}$ and $\gamma_{W} = \Gamma_{W}/m_{W}$. The three-body 
phase space is parametrized as follows \cite{Kuhn}:
\begin{eqnarray}
dR_{3} = \frac{1}{32}m^{2}_{t}~dx~dy~d(\cos\theta_{l})~d\alpha~d\beta,
\end{eqnarray}
By defining $x = \frac{2p_{1}.k_{1}}{m^{2}_{t}}$ and $y =
\frac{(k_{1}+k_{2})^{2}}{m^{2}_{t}}$ and from Fig.\ref{EventPlane}
one has:
\begin{eqnarray}\label{FV}
&&p_{1} = m_{t}(1;0,0,0)~,~k_{1} = \frac{m_{t}}{2}x(1;0,0,1)~,~k_{1}.s = -\frac{xm_{t}}{2}\cos\theta_{l},\\
\nonumber && p_{2} =
\frac{m_{t}}{2}(1-y)(1;\sin\theta_{b},0,\cos\theta_{b}),~k_{2} =
\frac{m_{t}}{2}(1-x+y)(1;-\sin\theta_{\nu},0,\cos\theta_{\nu}),
\end{eqnarray}
where
\begin{eqnarray}
\cos\theta_{\nu} = \frac{x(1+y-x)-2y}{x(1-x+y)}~,~\cos\theta_{b} =
\frac{2y-x-xy}{x(1-y)},
\end{eqnarray}
After replacing Eq.(\ref{FV}) in Eq.(\ref{MX}) and integration over two Euler 
angles and in the $W$ narrow width approximation, 
$\gamma_{W}\rightarrow 0$, the double differential $x-\theta_{l}$ distributions in the point-like 
four fermion limit is:
\begin{eqnarray}\label{DWidth}
&&\frac{d^{2}\Gamma}{dx~d\cos\theta_{l}}=\frac{G^{2}_{F}m^{3}_{t}m^{3}_{W}}{32\pi^{2}\Gamma_{W}}|V_{tb}|^{2}\times\nonumber\\
&&\left(F_{0}(x)+G_{0}(x)\cos\theta_{l}+\frac{1}{\Lambda^{4}_{NC}}[F_{1}(x)+G_{1}(x)
\cos\theta_{l}]\right),
\end{eqnarray}
where $\Lambda_{NC} = \frac{1}{\sqrt{|\overrightarrow{\theta}|^{2}}}$ is the scale of the noncommutativity
and $F_{0,1}(x)$ and $G_{0,1}(x)$ are defined in the following forms:
\begin{eqnarray}
F_{0}(x) &=& G_{0}(x) = x(1-x),\nonumber\\
F_{1}(x) &=& \frac{m^{4}_{W}(1-x)(r-x)(1-r)}{16x^{2}},\nonumber\\
G_{1}(x) &=& \frac{m^{4}_{W}(1-x)(r-x)(2r-x-rx)}{16x^{3}}.
\end{eqnarray}

\begin{figure}[ht]
\begin{center}
\epsfig{file=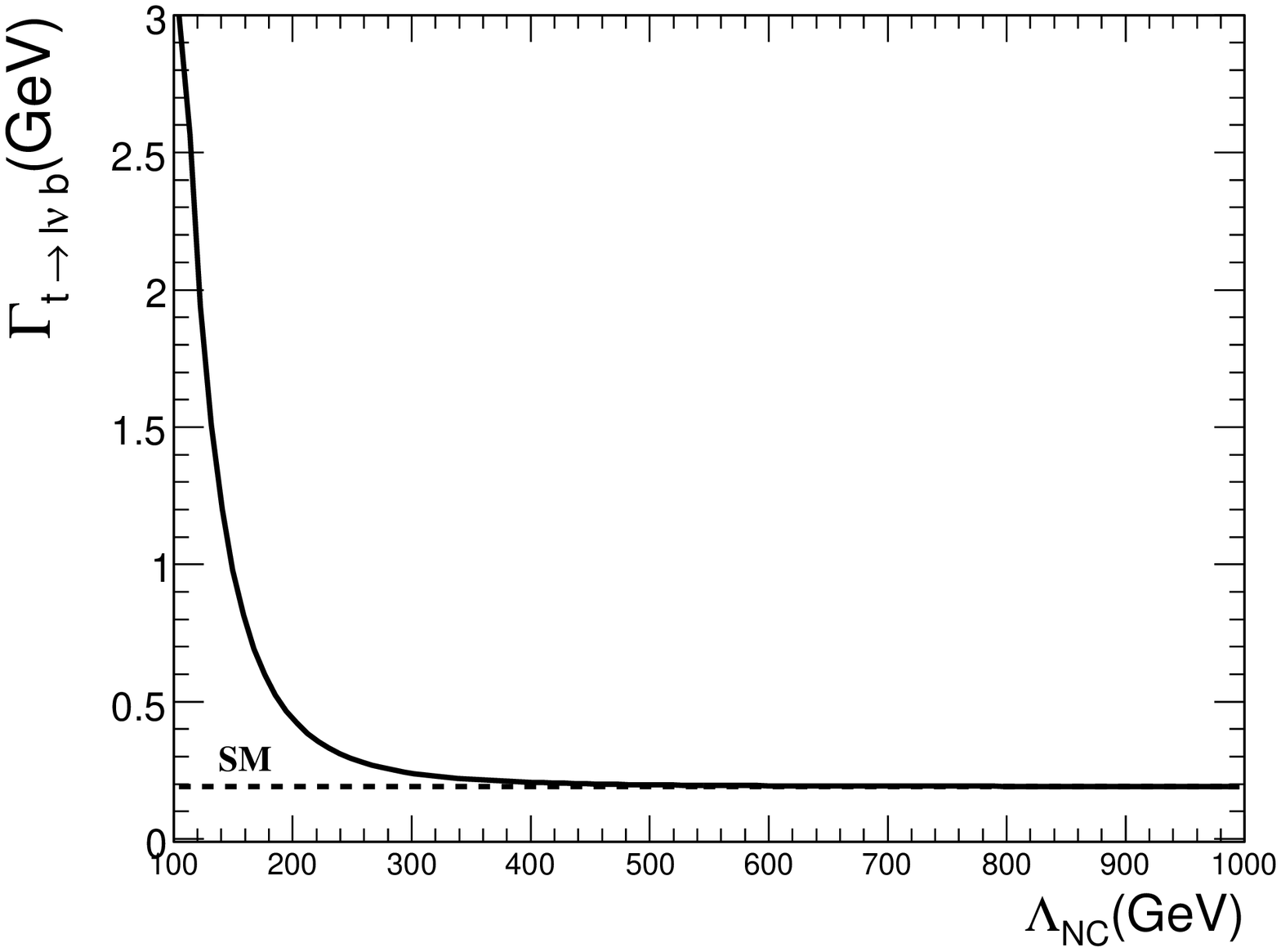,height=7cm,width=7cm}
\epsfig{file=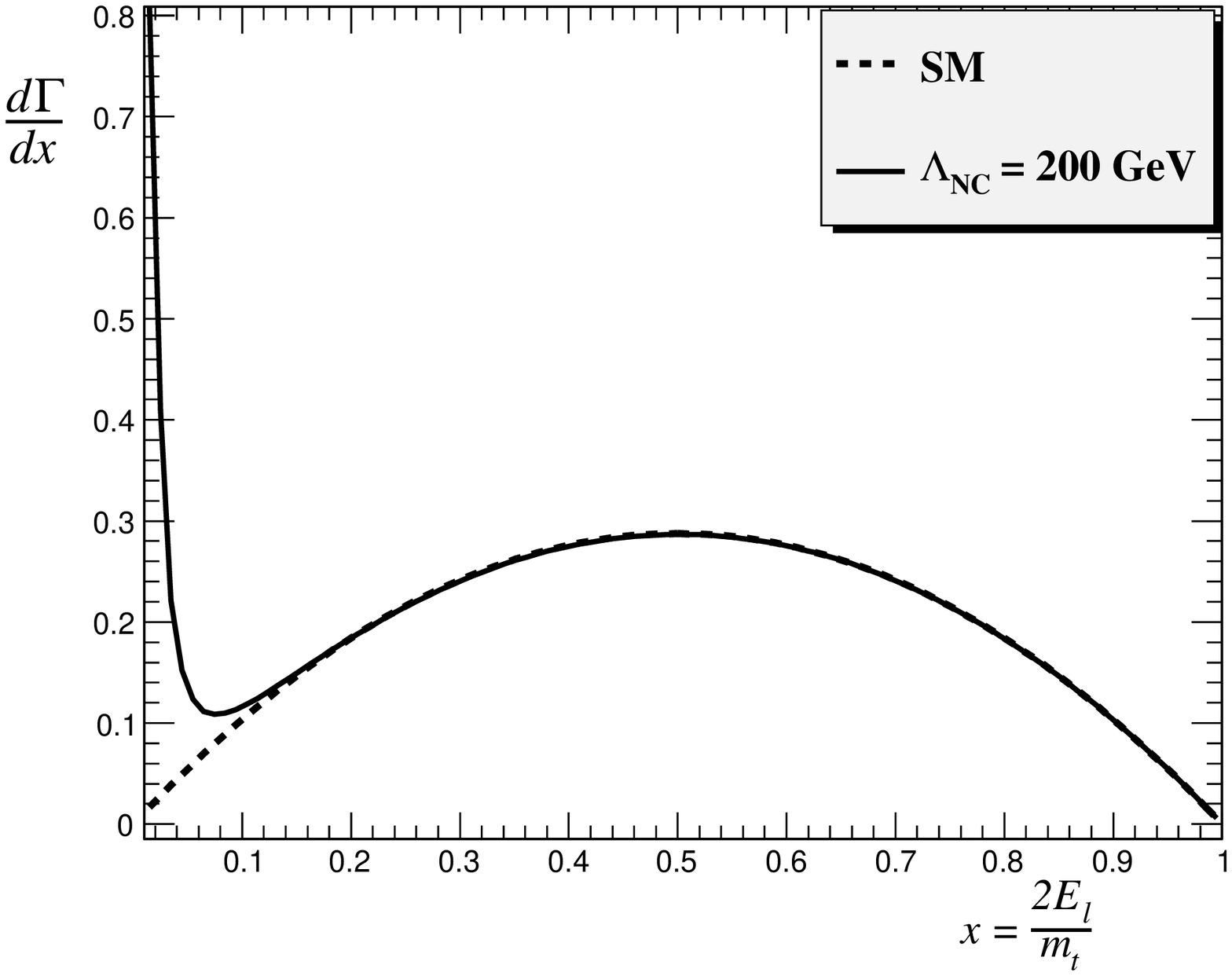,height=7cm,width=7cm}
 \caption{ Left: The three-body decay rate as a function of noncommutativity scale (solid curve). Right: 
  The energy spectrum (Born approximation) for the charged lepton in the top quark decay for
  the ordinary SM (dashed curve) and for when $\Lambda_{NC} = 200$ GeV in the NCSM (solid curve).}
    \label{fig:TopWidth}
\end{center}
\end{figure}
\begin{figure}[ht]
\begin{center}
\epsfig{file=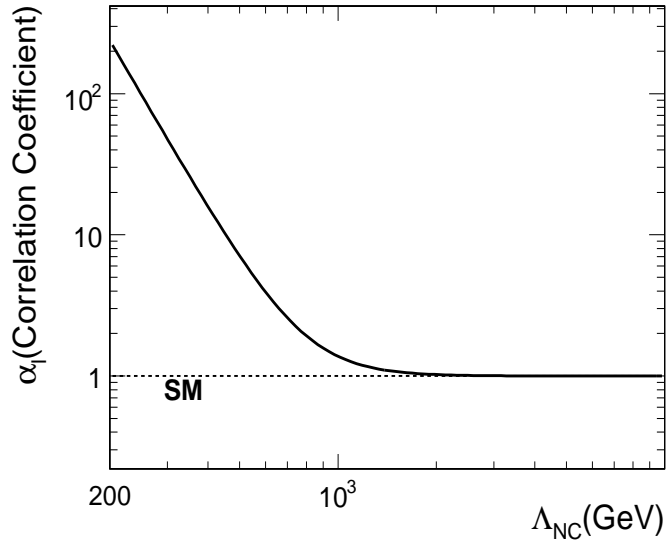,height=8cm,width=9cm}
 \caption{The Correlation Coefficient $\alpha_{l}$ as a function of noncommutativity scale in the logarithmic scale.}
    \label{fig:CorrCoef}
\end{center}
\end{figure}

In Fig.\ref{fig:TopWidth}, in the left the three-body top decay width is presented as a function of
noncommutativity scale. This figure shows that the noncommutative effect is 
negligible for $\Lambda_{NC}\geq 500$ GeV and is not observable.
The right plot in Fig.\ref{fig:TopWidth} shows the energy spectrum for
the charged lepton in the top quark decay at Born approximation for the ordinary SM (dashed curve)
and for when $\Lambda_{NC} = 200$ GeV in the NCSM (solid curve).
It shows that the probability of having very low energy charged leptons, $x < 0.1$ or $E_{l} < 9$ GeV,
is high because of the noncommutative effects. However, from the experimental point of view
this effect is not observable since we are restricted by the detector resolution. The detected
leptons with transverse momentum less than 15 GeV are usually  not reliable and rejected
in the experiments. 
It is noticeable that, this effect is present only for low values of noncommutativity scale.
The lepton energy spectrum is matched with the SM case for $\Lambda_{NC} \geq 500$ GeV.
\\
It is more useful to express the Eq.(\ref{DWidth}) in the following way:
\begin{eqnarray}\label{Dwidth1}
\frac{d^{2}\Gamma}{dx~d\cos\theta_{l}}=\frac{d\Gamma}{dx}\times\frac{1}{2}\left(1+\alpha_{l}(x)\cos\theta_{l}\right),
\end{eqnarray}
where $\alpha_{l}(x)$, {\it Correlation Coefficient} or {\it Spin Analyzing Power}, is:
\begin{eqnarray}
\alpha_{l}(x) = \frac{G_{0}(x)+G_{1}(x)/\Lambda^{4}_{NC}}{F_{0}(x)+F_{1}(x)/\Lambda^{4}_{NC}}.
\end{eqnarray}
In the ordinary standard model and in the limit of vanishing the lepton masses, 
$\alpha_{l}$ is independent of $x$ and is equal to one \cite{Mahlon}. As a result of Eq.(\ref{Dwidth1}), 
in the NCSM the angular distribution of a polarized top decay has the same form as
the ordinary SM:
\begin{eqnarray}
\frac{1}{\Gamma}\frac{d\Gamma}{d\cos\theta_{l}} = \frac{1}{2}(1+\alpha_{l}\vec{\bf \sigma}.\bf\hat{ p}_{l}),
\end{eqnarray}
where $\bf \hat{p}_{l}$ describes the direction of the flight of lepton in the rest frame of the top
quark and $\sigma_{i}$ denots the Pauli matrices. However, according to Fig.\ref{fig:CorrCoef} in the
NCSM the spin analyzing power, $\alpha_{l}$, depends on the noncommutativity scale.
Fig.\ref{fig:CorrCoef} reveals a significant deviation from the SM even for the case that 
$\Lambda_{NC}$ is around 1 TeV.

It is well known that top quarks produced in hadron colliders are scarcely polarized. As a result,
the correlations between top spin and anti-top spin in the $t\bar{t}$ is considered \cite{Beneke,Brandenburg}.
In order to illustrate we consider the dileptonic decay of the $t\bar{t}$ events in hadron colliders:
\begin{eqnarray}
\bf {PP,P\bar{P}} \rightarrow t\bar{t} + X \rightarrow l^{+}{l^{\prime-}} + X,
\end{eqnarray}
The double differential angular distribution of the leptons coming form the top and anti-top
in the ordinary SM is \cite{Beneke,Brandenburg}:
\begin{eqnarray}\label{spintt}
\frac{1}{\sigma}\frac{d^{2}\sigma}{d\cos\theta_{l^{+}}d\cos\theta_{l^{-}}} = \frac{1}{4}(1+
\kappa \cos\theta_{l^{+}}\cos\theta_{l^{-}})~,~\kappa = \alpha_{l^{+}}\alpha_{l^{-}}
\times\frac{N_{\|}-N_{\times}}{N_{\|}+N_{\times}}.
\end{eqnarray}
where $\theta_{l^{+}}$$(\theta_{l^{-}})$ is the angle between the direction of the $l^{+}(l^{-})$  
in the rest frame of the $t(\bar{t})$ and the $t(\bar{t})$ direction in the $t\bar{t}$ center of mass.
$N_{\|}$ is the number of top
pair events where both quarks have spin up or spin down
and $N_{\times}$ is the number of top pair events where one quark is spin up and the other is spin down.
Eq.(\ref{spintt}) shows the strong dependence of the experimantal observable, $\kappa$, to the spin analyzing 
power ($\alpha_{l}$).
For the $t\bar{t}$ production at the Tevatron, the SM predicts $(N_{\|}-N_{\times})/(N_{\|}+N_{\times}) = 0.88$. 
The D{\O} measurement for the quantity $\kappa$ is $2.3\pm 2.5$ \cite{D0}, which is not very good
due to low statistics.
Having a good measurement for the $\kappa$ and knowing the depenedency of $N_{\|}$ and $N_{\times}$ on the
noncommutativity scale might provide valuable information concerning the noncommutativity scale.
Just as an example, using this rough measurement from D{\O} and ignoring of
any dependency of $N_{\|}$ and $N_{\times}$ on
the noncommutativity scale yields $\Lambda_{NC} \sim 900$ GeV.

\section{Conclusion}
The noncommutative effects only show up for the low values of the noncommutativity scale in
the lepton energy spectrum from the decay of top quarks. The effects are not visible for low values of
the noncommutativity scale because of the experimental restrictions.
However, the charged lepton spin correlation coefficient is sensitive to the noncommutativity scale
from very low values to 1 TeV. It is a powerful tool for investigation of non-SM effects, particularly
in $t\bar{t}$ events.

\section{Acknowledgment}
The author thanks to Fabrice Hubaut for the useful discussions.

\end{document}